\documentclass[draft]{agujournal2019}
\usepackage{url} 
\usepackage{lineno}
\usepackage[finalnew]{trackchanges}
\usepackage{soul}

\draftfalse

\journalname{Geophysical Research Letters}

\begin{document}

%
%


\title{Forecasting the Future with Yesterday’s Climate: Temperature Bias in AI Weather and Climate Models}

%
%




\authors{Jacob B. Landsberg$^{1,2}$, Elizabeth A. Barnes$^{1,2,3}$}

\affiliation{1}{Department of Atmospheric Science, Colorado State University, Fort Collins, CO, USA}
\affiliation{2}{Faculty of Computing and Data Sciences, Boston University, Boston, MA, USA}
\affiliation{3}{Department of Earth and Environment, Boston University, Boston, MA, USA}





\correspondingauthor{Jacob B. Landsberg}{jlandsbe@bu.edu}



\begin{keypoints}
\item AI weather and AI climate models produce cold-biased boreal winter land temperatures that resemble those from 15-20 years earlier.
\item The weather model cold bias is strongest for the hottest temperatures, suggestive of limited training exposure to modern extreme heat.
\item The climate model cold bias is largest in regions, seasons, and parts of the temperature distribution where climate change has been largest.
\end{keypoints}
                                              
%
%

%
%


\begin{abstract}

AI-based climate and weather models provide fast, skillful forecasts yet face a key challenge: predicting future climates while being trained with historical data. We investigate this issue by analyzing boreal winter land temperature biases in AI weather (FourCastNet V2 Small and Pangu Weather) and climate (Ai2 Climate Emulator version 2) models. We evaluate these models during time periods that are significantly more recent than the bulk of their training data, allowing us to assess how well they generalize to more modern, conditions. We find that all models produce cold-biased mean temperatures, resembling climates from 15–20 years earlier than their prediction period. Furthermore, FourCastNet’s and Pangu's cold bias is strongest for the hottest predicted temperatures, indicating limited training exposure to modern extreme heat events. In contrast, ACE2’s bias is more evenly distributed but largest in regions, seasons, and parts of the temperature distribution where historic global warming is most pronounced.

\end{abstract}

\section*{Plain Language Summary}
AI-based climate and weather models, which learn from historical data, can struggle to accurately predict future conditions, especially as the climate changes. We probe this issue by analyzing boreal winter land temperature biases in AI weather models (FourCastNet and Pangu) and an AI climate model (ACE2). We find that all models produce temperatures that better resemble climates from 15–20 years earlier than the period they are predicting. In some regions, like the Eastern U.S., the predictions resemble climates from as much as 20–30 years earlier. Further analysis shows that FourCastNet’s cold bias is strongest in the hottest predicted temperatures, indicating that these models may not have seen enough examples of modern extreme heat events in the past data. In contrast, ACE2’s bias is more evenly distributed but largest in regions, seasons, and parts of the temperature distribution where climate change has been most pronounced. These findings underscore the challenge of training AI models exclusively on historical data and highlight the need to account for such biases when applying them to future climate prediction.

\newpage

%
%

%


%
%
%
%
\section{Introduction}
Over the last five years, a new generation of fully data-driven AI models has emerged, reimagining weather forecasts and exploring early applications to climate prediction \cite{atmos16010082}. Unlike traditional dynamical models, which are governed by physical equations, these AI models learn relationships between variables directly from large datasets (e.g., \citeA{EbertUphoffHilburn2023,Rasp2020, bonev2023sphericalfourierneuraloperators}). This approach has largely been successful, with many recent AI models achieving state-of-the-art performance (e.g, \citeA{pathak2022fourcastnetglobaldatadrivenhighresolution,bonev2023sphericalfourierneuraloperators, bonev2025fourcastnet3geometricapproach,Bi2023,Lam2023,lang2024aifsecmwfsdatadriven}). Furthermore, these models are much less computationally expensive than dyanmical models, allowing for faster predictions and more extensive ensemble simulations \cite{Liu2024}. 

One of the key challenges with fully data-driven AI models is that they are most often trained on historical data, which may not accurately represent future conditions. This can lead to biases in the model's predictions, particularly in the context of a changing climate \cite{lindsey2020climate}. For example, these models may be tasked with predicting temperatures that largely lie outside the bulk of their training distribution \cite{Beucler2024climate}. \citeA{rackow2024robustnessaibasedweatherforecasts} examined this phenomenon by assessing the performance of three prominent AI weather models: Pangu Weather (Pangu) \cite{Bi2023}, Graphcast \cite{Lam2023}, and the AIFS \cite{lang2024aifsecmwfsdatadriven} under different climate regimes. The former two models were trained on reanalysis data from 1979-2017 \cite{Bi2023,Lam2023}, while the later was trained on data from 1979-2020 \cite{lang2024aifsecmwfsdatadriven}. \citeA{rackow2024robustnessaibasedweatherforecasts} confronted these models with a preindustrial climate (1955), a modern climate (2023), and a future, warmer climate (2049). They found that for these three different years, while the models' biases varied, in general, they exhibited warmer biases in the preindustrial climate, slight cold biases in the modern climate, and significant cold biases in the future climate. Furthermore, \citeA{zhang2025numericalmodelsoutperformai,Kent2025} both find that AI weather and AI climate models perform worse than traditional models when predicting record-breaking events. These works, therefore, underscore the challenges of training AI models on historical data.

In this study, we analyze boreal winter land temperature predictions of two AI weather models and an AI climate model with explicit CO$_2$ forcing more broadly. Specifically, we quantify the extent to which FourCastNet V2 small (FourCastNet) \cite{pathak2022fourcastnetglobaldatadrivenhighresolution}, Pangu \cite{Bi2023}, and Ai2 Climate Emulator version 2 (ACE2) \cite{WattMeyer2025} reflect their evaluation-period climate as opposed the mean climate of their training data.  For FourCastNet and Pangu, we focus on predictions from 2020–2025, while for ACE2 we analyze those from 1996–2010. These periods are outside of their respective training sets and are also warmer than their training climatologies. In doing so, we assess the persistence of training-climate influence on the models’ temperature distributions. Finally, we probe when and where these biases are most pronounced across the different types of models.

\section{Data and Models}
\subsection{FourCastNet and Pangu}
FourCastNet \cite{bonev2023sphericalfourierneuraloperators} and Pangu \cite{Bi2023} are both fully-data-driven AI weather models designed by NVIDIA and Huawei, respectively. Pangu is trained with ECMWF Reanalysis v5 (ERA5) data \cite{Hersbach:2020aa} from 1979-2017 and a transformer architecture, while FourCastNet utilizes ERA5 data from 1979-2015 and a Spherical Fourier Neural Operator (SFNO) architecture. Thus, the training data for FourCastNet and Pangu are centered around 1997 and 1998, respectively.  We use both models' 2m temperature (2mT) outputs generated by \citeA{AcceleratingCommunityWideEvaluationofAIModelsforGlobalWeatherPredictionbyFacilitatingAccesstoModelOutput} for 2-day and 9-day leads.\remove{These forecasts are initialized with 0000 UTC NOAA Global Forecasting System (GFS) data \protect{\cite{noaa_gfs_aws}}} \remove{to make predictions using 6-hour timesteps over the 5-year December-January-February (DJF) period between December 2020 and February 2025.}\add{Each forecast is initialized with 0000 UTC NOAA Global Forecasting System (GFS) data \protect{\cite{noaa_gfs_aws}}}\add{ from 2 or 9 days prior to the forecast date. Predictions are rolled out in 6-hour timesteps during the December-January-February (DJF) period between December 2020 and February 2025.} As the training sets for FourCastNet and Pangu are centered around the turn of the century, this timeframe is not only more modern than any training year, it is also $\sim25$ years more modern than the average training year. We compute the daily average of these 6-hour forecasts to obtain daily mean temperature forecasts. Both models are missing a small number of initialization dates, although this represents less than 0.7\% of the total data for FourCastNet and less than 0.9\% of the data for Pangu. For more details see \ref{sec:missing_data}.

\subsection{ACE2}
ACE2 is an atmosphere-only AI climate model designed to produce stable $\sim\!100$-year simulations of Earth's climate, capturing atmospheric variability, global temperature trends, and tropical phenomena like the Madden Julian Oscillation and El Niño Southern Oscillation \cite{WattMeyer2025}. ACE2 similarly uses an SFNO architecture, a 6-hour autoregressive structure, and is trained on ERA5 data from 1940–1995, 2011–2019, and 2021-2022. Unlike FourCastNet, ACE2 includes specific CO$_2$ forcing from the Climate Model Intercomparison Project – Phase 6  \cite{gmd-10-2057-2017} and NOAA Global Monitoring Laboratory \cite{NOAA_CO2}. We generate a 5-member ensemble of daily surface temperatures from 1940-2020. The first ensemble member is generated by running the model with January 1940 initial conditions from ERA5. For each subsequent ensemble member, we initialize the model using the day-1 forecast from the previous ensemble member as its initial condition. \add{All ensemble members use the same sea surface temperature and CO$_2$ forcings and differ only in their initializations.} We then compute the daily mean temperature for each ensemble member, and extract data between 1996-2010. \add{While none of these years are within ACE2's training data, we do include 1996-2000, which are in the validation set. We find, including or excluding these years has minimal impact on the global bias (-.33 K compared to -.35 K). }We chose this earlier period, rather than the 2020-2025 time range we use for the weather models, as ACE2 includes training data up through 2022. Nontheless, as with the weather models, 1996-2010 is still 25-30 years more modern than the average training year ($\sim1975$).

\subsection{Temperature Comparison}
To evaluate the daily temperature biases of the models, we subset the data to boreal winter land temperatures, which we define to be data in the DJF period and all land gridpoints except those in Antarctica and Greenland. We focus on boreal winter, as both cold extremes in the Northern Hemisphere and land temperatures in particular have been shown to be warming more rapidly than the global average \cite{esd-11-97-2020,crimmins2023fifth}. We analyze the biases of these models by comparing their predictions to \add{daily }ERA5\add{ formed by sampling at 6-hour frequency}—the dataset with which all three models were trained. We use ERA5 data at 0.25° resolution\add{, which we then coarsen to 1°,} when comparing with FourCastNet and Pangu and \add{directly} at 1° resolution when comparing to ACE2, as those are the native resolutions of the respective AI models. To compute bias, we take the time-mean difference at every grid point between ERA5 and the model predictions. Global mean biases are then reported as the cosine-latitude-weighted average of these gridpoint biases. \add{To compute the spatial significance of these biases, we use a two-sided Z-test. We compute the population variance by subtracting the seasonal mean temperature from ERA5 and computing variance from 1980-2025 for the weather models and from 1940-2022 for ACE2. We perform this test with the null hypothesis that the mean bias is equal to zero at an $\alpha = 0.1$. We use the lag-1 autocorrelation to estimate the effective sample size when computing the Z-statistic for all models (Equation 6 in \protect{\citeA{Santer2000}}).}
\section{Results}

\subsection{Weather Models}
\begin{figure}[htbp!]
    \includegraphics[width=\textwidth]{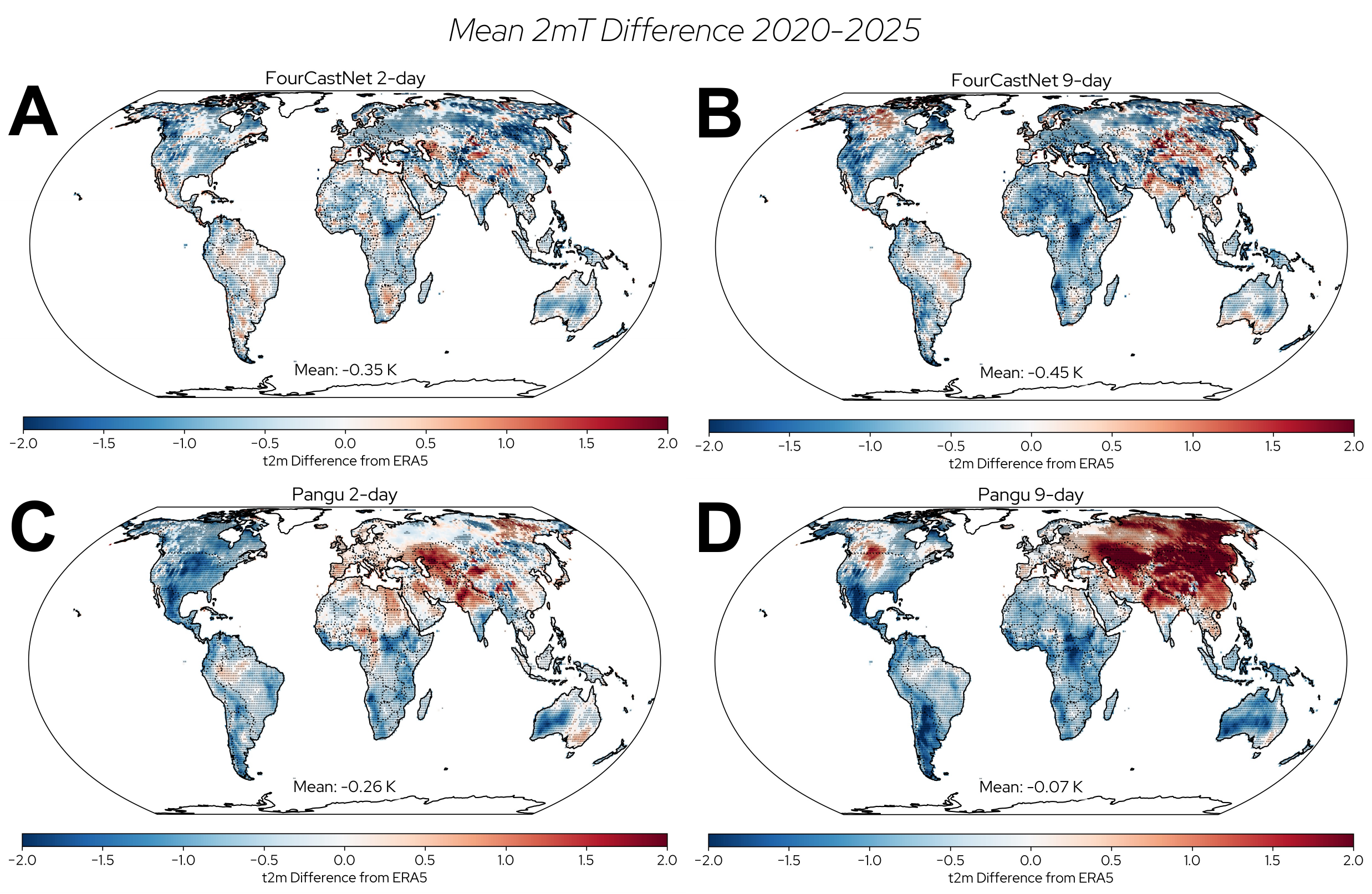}
    \caption{
        Mean 2mT differences for 2020-2025 boreal winter land temperatures compared to ERA5 for (a) FourCastNet 2-day lead, (b) FourCastNet 9-day lead, (c) Pangu 2-day lead, and (d) Pangu 9-day lead. Global means are shown at the bottom of each panel\add{, with stippling indicating statistically significant non-zero bias (see Methods)}.
    }
    \label{fig:figure1_fourcast_pangu_cold_bias}
\end{figure}

We find that at 2-day and 9-day lead times, FourCastNet and Pangu both produce forecasts of 2020-2025 boreal winter land temperatures that are too cold relative to ERA5 (Figure \ref{fig:figure1_fourcast_pangu_cold_bias})\add{, a pattern that holds across lead times (Figure S1)}. While both models are cold, FourCastNet is colder than Pangu with global mean differences of -0.35 K  and -0.45 K compared to -.26 K and -.07 K for 2-day and 9-day leads respectively. Moreover, this cold bias is distributed nearly globally, with the exception of Asia in Pangu's 9-day lead forecasts.

\begin{figure}
    \centering
    \includegraphics[width=\textwidth]{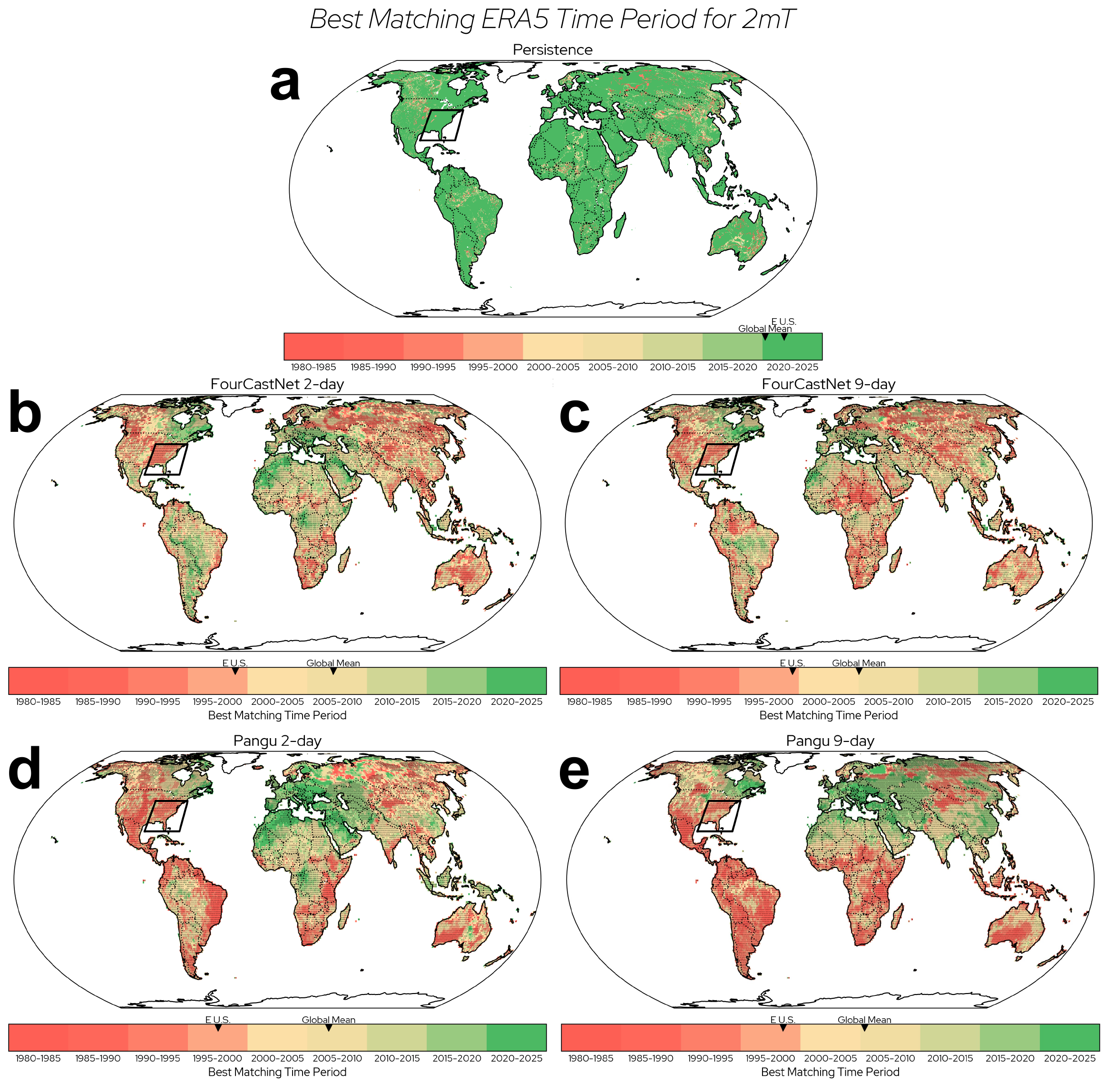}
    \caption{The closest matching 5-year span of ERA5 land temperatures to FourCastNet and Pangu's 9-day lead forecasts of 2020-2025 boreal winter land temperatures for a) a 9-day persistence forecast, b) FourCastNet 2-day prediction, c) FourCastNet 9-day prediction, d) Pangu 2-day prediction, and e) Pangu 9-day prediction. The Eastern U.S. (highlighted by the black box) and global mean time period are shown in the legend.\add{ Stippling indicates grid points that have statistically significant non-zero bias.}}
    \label{fig:figure2_fourcast_pangu_lagging_climate}
\end{figure}

In fact, the temperatures generated by FourCastNet and Pangu for 2020-2025 more closely resemble temperatures from 15-20 years earlier (Figure \ref{fig:figure2_fourcast_pangu_lagging_climate}b-e). For some regions, like the Eastern U.S. (25°N to 42°N and 70°W to 95°W), this bias is even more pronounced, with the model's forecasts most closely resembling ERA5 temperatures from 20-25 years earlier. This suggests these models may be struggling to fully generalize to 2020-2025 which lies beyond their training data's climate, which is centered about 25 years prior. We compare these models' lagging climate to that generated by a 9-day persistence forecast (Figure \ref{fig:figure2_fourcast_pangu_lagging_climate}a). We find that a persistence forecast shows essentially matching mean temperatures to the prediction period of 2020-2025. While this is perhaps unsurprising, as \change{a 9-day persistence forecast shares roughly 90\% of its data with the ERA5 truth data}{most of the ERA5 persistence forecasts are the same data as the ground truth ERA5 data}, it highlights that a much simpler prediction model can offer a more temporally consistent mean climate than FourCastNet and Pangu.

\subsection{Extreme Modern Temperatures}
We further investigate where this difference in modern temperatures is most pronounced by looking at the tails of the temperature distribution for the 9-day lead time forecasts. We find that the hottest temperature forecasts for both Pangu and FourCastNet exhibit a much stronger cold bias than those for the coldest temperatures. For instance, the hottest 10\% of 2020--2025 temperature forecasts are on average 0.91 K colder than ERA5 for FourCastNet and 0.34 K colder for Pangu (Figures \ref{fig:figure3_fourcast_pangu_extreme_temperatures}b and \ref{fig:figure3_fourcast_pangu_extreme_temperatures}d), while the coldest 10\% of FourCastNet's temperatures exhibit minimal bias compared to ERA5 and Pangu's are even 0.12 K warmer (Figures \ref{fig:figure3_fourcast_pangu_extreme_temperatures}a and \ref{fig:figure3_fourcast_pangu_extreme_temperatures}c). An example of the temperature distributions' tail behavior for the SE U.S (30°–35° N, 90°–100° W)
 is shown in \ref{fig:figure3_fourcast_pangu_extreme_temperatures}e, with Pangu and FourCastNet both matching much more closely with ERA5 for the cold tail than the hot tail. 
 
 \begin{figure}
    \centering
    \includegraphics[width=\textwidth]{ACE_fig3_v2.pdf}
    \caption{Mean 2mT differences as in Figure \ref{fig:figure1_fourcast_pangu_cold_bias} but for the 10th and 90th percentiles of FourCastNet's (a, b) and Pangu's (c,d) 9-day lead forecasts\add{, with stippling indicating statistically significant non-zero bias}. An example of the tail behavior for the SE U.S. (bounded by the yellow box in a-d) is shown in e. The global mean percent of training data as or more extreme than the 10th and 90th percentiles of 2020-2025 ERA5 temperatures is displayed in f.}
    \label{fig:figure3_fourcast_pangu_extreme_temperatures}
\end{figure}

 This stark difference in bias between the coldest and hottest temperatures may be a reflection of the models' training data, which is primarily from a colder climate. For instance, globally there is $\sim 2$-$3\times$ as much training data that is as cold or colder than the 10th percentile of 2020-2025 ERA5 temperatures than there is training data that is as hot or hotter than the 90th percentile of ERA5 temperatures (Figures \ref{fig:figure3_fourcast_pangu_extreme_temperatures}f and S2). Hence, this cold bias during the hotter forecasts is likely a pull toward the mean of the training dataset.  These findings hold for various percentile thresholds used to define cold and hot tails (Figures S3 and S4).

\subsection{Climate Models}
\begin{figure}
    \centering
    \includegraphics[width=\textwidth]{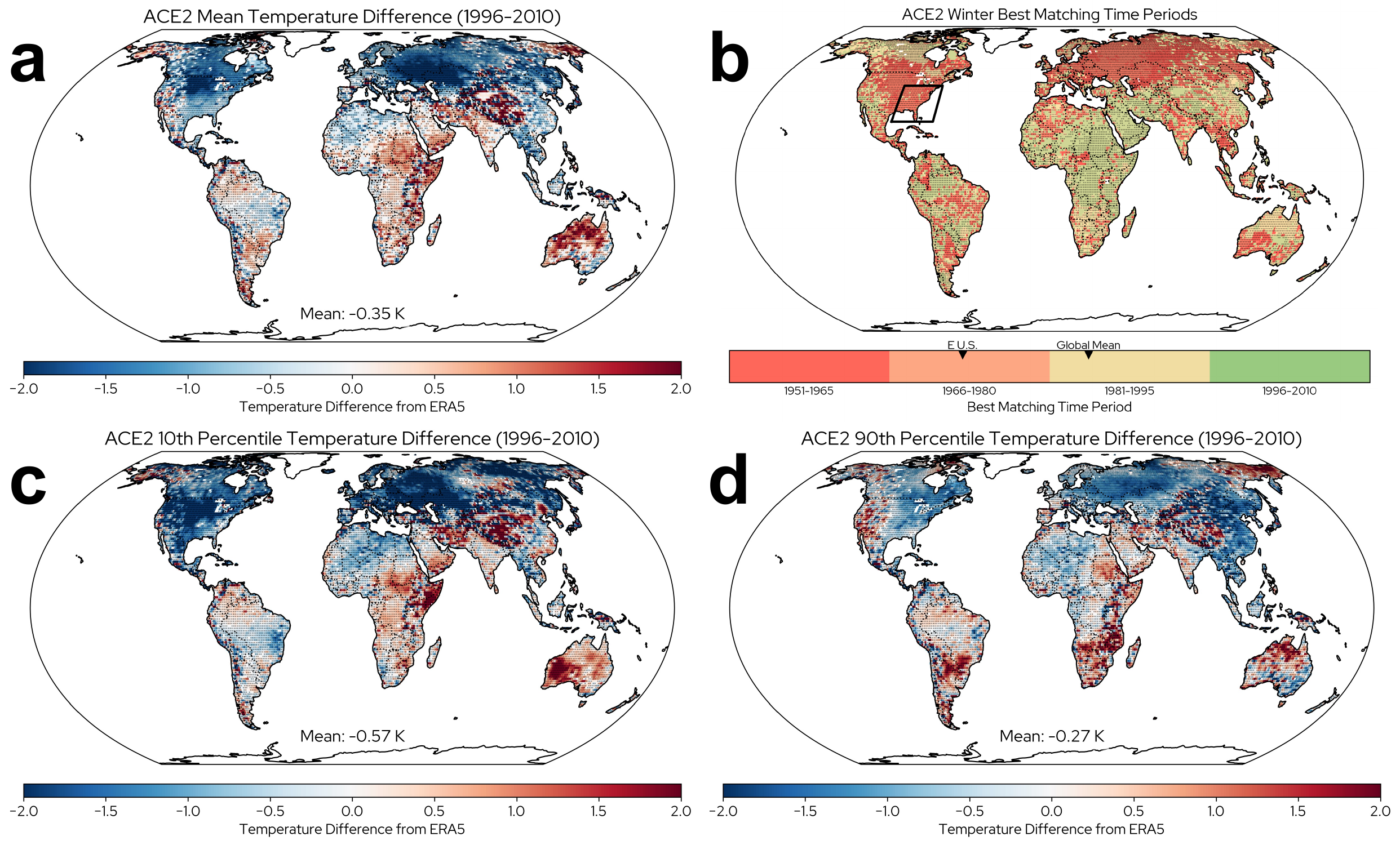}

    \caption{a) Mean surface temperature differences for 1996-2010 boreal winter land temperatures compared to ERA5 for ACE2. b) The closest matching 15-year span of ERA5 land temperatures to ACE2's 1996-2010 boreal winter land temperatures. The Eastern U.S. (highlighted by the black box) and global mean time period are shown in the legend. c) Mean surface temperature differences as in (A) but for the 10th percentile of ACE2's 1996-2010 predictions. d) Mean surface temperature differences as in (a) but for the 90th percentile of ACE2's 1996-2010 predictions. Global means are shown at the bottom of (a), (c), and (d)\add{, while all figures show stippling at grid points of statistically significant non-zero bias}.}
    \label{fig:figure4_ace2_bias}
\end{figure}

We similarly analyze ACE2 temperatures, investigating how a climate model with an SFNO architecture, like FourCastNet, but adapted for long-term climate prediction through the inclusion of CO$_2$ forcing, performs.  We find that ACE2 is also too cold, with a global mean of -.35 K relative to ERA5, and is particularly cold over North America, Europe, and Russia (Figure \ref{fig:figure4_ace2_bias}a). \add{This cold bias represents 35\% of the average root mean square error compared to ERA5 over the five ensemble members.} As with the weather models, this connotes a temperature pattern more similar to that of 15-20 years prior, with some regions, like the Eastern U.S. lagging $\sim30$ years behind (Figure \ref{fig:figure4_ace2_bias}b). This again is consistent with a pull towards the mean climate of the training data, which is centered around 1975. However, unlike FourCastNet and Pangu\change{,  ACE2's bias is not strongly concentrated in the hottest temperatures, but actually is further from ERA5's temperatures in the cold tail of the temperature distribution}{, which showed a stronger cold bias for their hottest forecasts, ACE2 exhibits a larger cold bias for its coldest predictions} (Figures \ref{fig:figure4_ace2_bias}c and \ref{fig:figure4_ace2_bias}d).

\begin{figure}
    \centering
    \includegraphics[width=\textwidth]{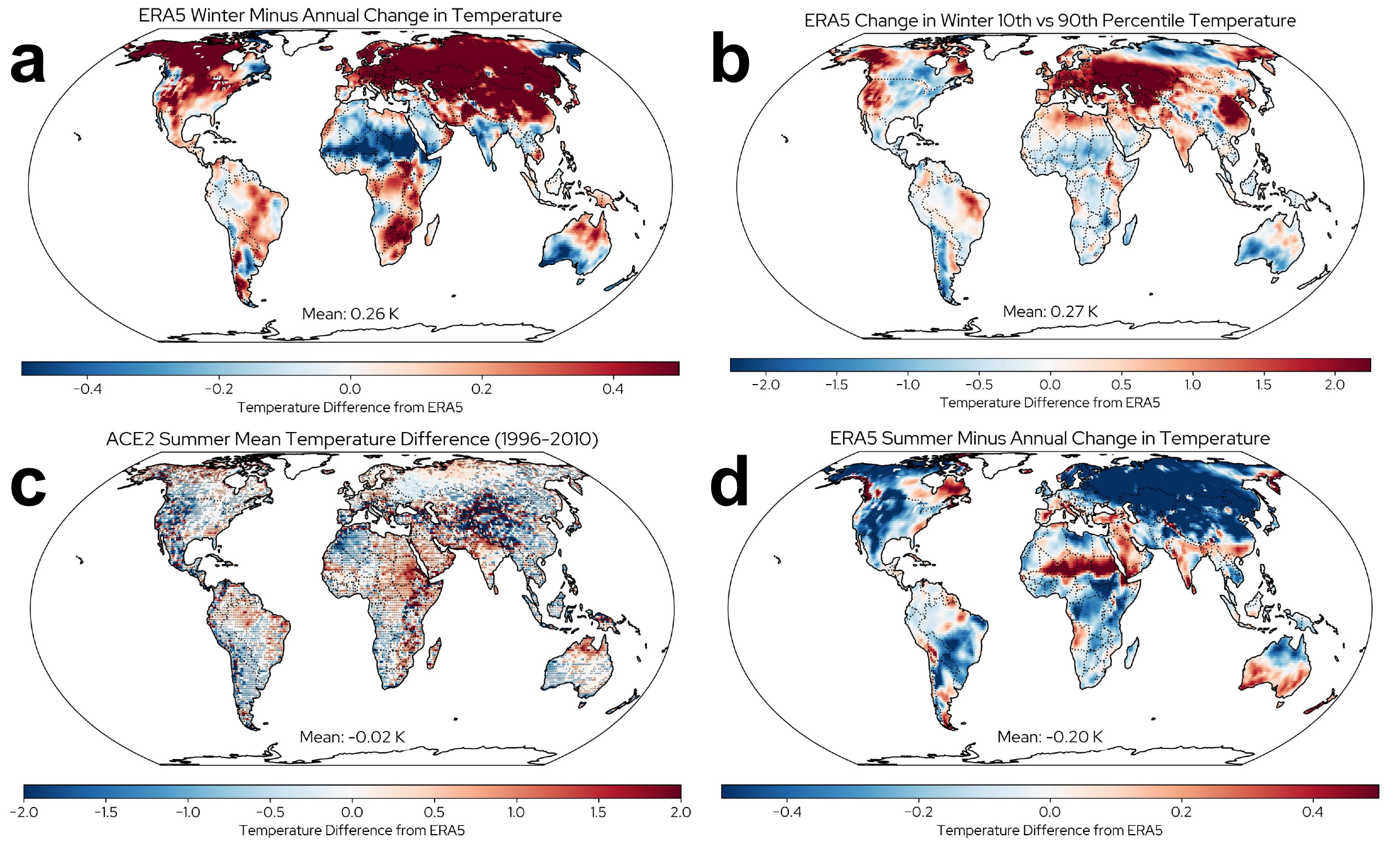}

    \caption{a) Change boreal winter land surface temperatures between 1940-1979 and 1980-2022 relative to the annual mean change. b) Change in the 10th vs. 90th percentile of boreal winter land surface temperatures between 1940-1979 and 1980-2022. c) Mean surface temperature differences for 1996-2010 boreal summer land temperatures compared to ERA5 for ACE2\add{. Stipping indicates statistically significant non-zero bias}. d) Change in boreal summer land surface temperatures between 1940-1979 and 1980-2022. Global means are shown at the bottom of each panel.}
    \label{fig:figure5_ace2_climate_change}
\end{figure}

We further analyze this asymmetry by situating ACE2’s biases in the context of climate change. To estimate the effect of climatic warming, we compute the difference between temperatures from 1980–2022 and those from 1940–1979. In line with previous work, (e.g. \citeA{esd-11-97-2020}), we find winter temperatures have warmed more rapidly than the annual mean, particularly over North America, Europe, and Russia (Figure \ref{fig:figure5_ace2_climate_change}a)—the same regions where ACE2 is most cold-biased (Figure \ref{fig:figure4_ace2_bias}a). Similarly, when we look at the change in the 90th percentile of winter temperatures compared to the 10th percentile, we find that the coldest winter temperatures have warmed more rapidly than the hottest temperatures over much of the Northern Hemisphere (Figure \ref{fig:figure5_ace2_climate_change}b). Again, this mirrors the stronger bias ACE2 shows in the cold tail (Figures \ref{fig:figure4_ace2_bias}c-d). Notably, the weather models, which show the opposite tail behavior to ACE2, have training periods that lack a similar asymmetric climatic warming trend (Figure S5). This pattern is consistent across seasons as well; for example, in boreal summer, ACE2 exhibits lower bias (Figure \ref{fig:figure5_ace2_climate_change}c), consistent with the fact that summer temperatures have warmed less rapidly than both winter and the annual mean (Figure \ref{fig:figure5_ace2_climate_change}d). This shows that ACE2's bias is largest in regions, seasons, and parts of the temperature distribution where climate change has been most pronounced.

\section{Discussion and Conclusions}
In this work we have shown that both AI weather and climate models exhibit cold biases when predicting  modern climates that lie outside of the bulk of their training data (Figures \ref{fig:figure1_fourcast_pangu_cold_bias} and \ref{fig:figure4_ace2_bias}). Instead, their boreal winter land temperatures better resemble those of 15-20 years prior (Figures \ref{fig:figure2_fourcast_pangu_lagging_climate} and \ref{fig:figure4_ace2_bias}b). This is consistent with a pull toward the mean of their training data, as all models have training data centered $\sim25-30$ years prior to their prediction time period. 

We did, however, find that the tails of the AI weather and AI climate temperature distributions displayed different behavior. FourCastNet and Pangu exhibited a cold bias almost exclusively for the hottest temperature predictions (Figure \ref{fig:figure1_fourcast_pangu_cold_bias}), which may be due to a lack of training data for modern extreme heat events (Figures \ref{fig:figure3_fourcast_pangu_extreme_temperatures}f and S2). This finding is aligned with \citeA{zhang2025numericalmodelsoutperformai}, who found that AI weather models performed poorly when predicting record-breaking (i.e., outside of the training set) extremes. ACE2, on the other hand, exhibited a more pronounced cold bias for the coldest temperature predictions (Figures \ref{fig:figure4_ace2_bias}c-d). We attribute this asymmetric bias to the pattern of warming temperatures under climate change in ACE2's training set. \add{Moreover, we found a much more pronounced winter bias than summer bias, a seasonal pattern that may be more difficult to see when looking at annual means \protect{{\cite{WattMeyer2025}}}.} We show that these spatial, seasonal, and distributional patterns of ACE2's bias align well with regions of rapid historical warming (Figure \ref{fig:figure5_ace2_climate_change}). Thus, although AI weather and AI climate models have different bias patterns in the tails of their temperature distributions, both are consistent with an anchoring to their training sets. While continued work is needed to fully understand the training mechanisms behind these biases, our findings highlight that simply including CO$_2$ forcing in an AI model (e.g., as in ACE2) is not sufficient to fully eliminate training-set artifacts.

Our work contributes to the growing body of literature documenting the limitations of AI models in extrapolating to climates outside their training domain \cite{rackow2024robustnessaibasedweatherforecasts,zhang2025numericalmodelsoutperformai,Kent2025,Hernanz2022}. We show that biases in both AI weather and climate models are already evident in present-day predictions, not only in future climates, and that these biases vary across space, season, and the temperature distribution. Several strategies have been proposed to mitigate such biases, including augmenting training data with climate model simulations that extend into the future \cite{Clark2022} or transforming inputs to be ``climate invariant'' \cite{Beucler2024climate}. Advancing this focus on developing more climate-robust AI models is critical. Since many AI models already achieve skill comparable to traditional approaches (e.g., \citeA{pathak2022fourcastnetglobaldatadrivenhighresolution,bonev2023sphericalfourierneuraloperators,bonev2025fourcastnet3geometricapproach,Bi2023,Lam2023,lang2024aifsecmwfsdatadriven}), addressing these biases will further strengthen their value for predicting both present and future climate.

\appendix 

\section{Missing Data}
\label{sec:missing_data}
Three initialization dates are missing from \citeA{AcceleratingCommunityWideEvaluationofAIModelsforGlobalWeatherPredictionbyFacilitatingAccesstoModelOutput} FourCastNet's run: December 4th, 2021, December 1st, 2024, and January 22nd, 2025. Hence, for 2-day lead forecasts, December 6th, 2021; December 3rd, 2024; January 24th, 2025 and for 9-day lead forecasts, December 13th, 2021, December 10th, 2024, and January 31st, 2025 are excluded. These three missing dates represents only 0.67\% of the total daily data we utilize from ERA5. Similarly, Pangu is missing: December 4th, 2021, December 1st, 2024, December 11th, 2024, and January 2nd, 2025. These four missing dates represent 0.89\% of the total daily data we utilize from ERA5.
\clearpage

\section*{Open Research Section}
ERA5 data are available from \cite{Hersbach:2020aa} at \url{https://doi.org/10.24381/cds.4991cf48}. FourCastNet and Pangu data are available from \citeA{AcceleratingCommunityWideEvaluationofAIModelsforGlobalWeatherPredictionbyFacilitatingAccesstoModelOutput}. The ACE2 model checkpoint and forcings are available from \citeA{WattMeyer2025} at \url{https://doi.org/10.57967/hf/5377}. All code and initialization files for ACE are available at \url{https://github.com/jlandsbe/AI_Bias}, which will be given a permanent DOI via Zenodo upon publication.

\section*{Conflict of Interest declaration}
The authors declare there are no conflicts of interest for this manuscript.

\acknowledgments
This work is funded, in part, by NOAA grant \#NA22OAR4310621 and Heising-Simons
Foundation grant \#2023-4720. The authors would also like to thank the two anonymous reviewers for their helpful comments and suggestions.

\clearpage

\bibliography{AI_biases}

\end{document}


\maketitle

\section*{Contents of this file}
Figures S1 to S5

\section*{Introduction}
Here we include 5 additional figures referenced in the main text. All figures are generated using the same data and methods as described in the main text.

\clearpage

\begin{figure}
    \centering
    \includegraphics[width=0.99\textwidth]{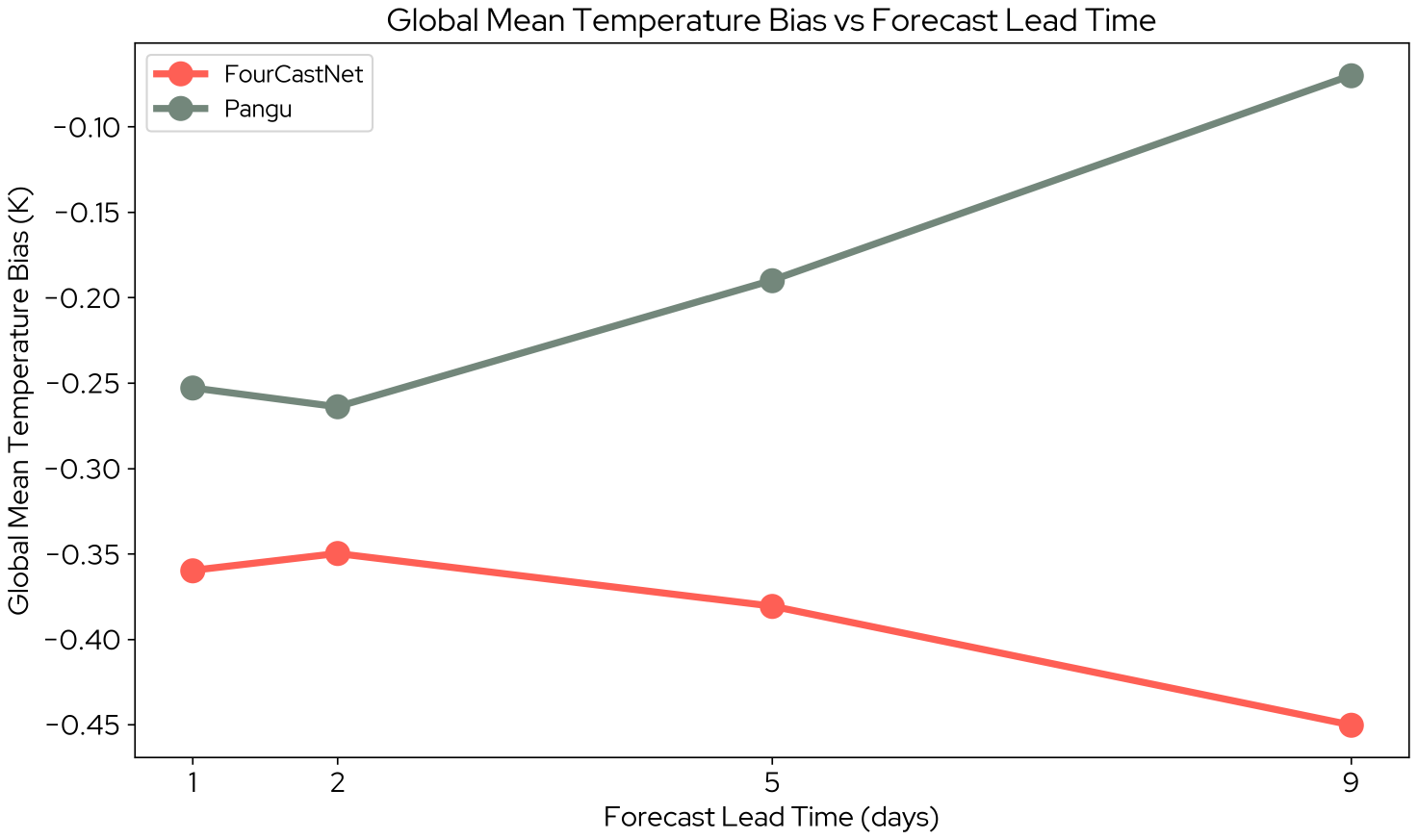}
    \caption{\add{Global temperature bias at 1-day, 2-day, 5-day, and 9-day lead times for FourCastNet and Pangu. Even at 1-day lead, both models exhibit a cold bias, with the bias generally increasing with lead time for FourCastNet and decreasing with lead time for Pangu (although as shown in Figure 1, there is still significant, offsetting regional bias at longer leads).}}
    \label{fig:figureS0_fourcast_pangu_cold_bias}
\end{figure}

\begin{figure}
    \centering
    \includegraphics[width=0.99\textwidth]{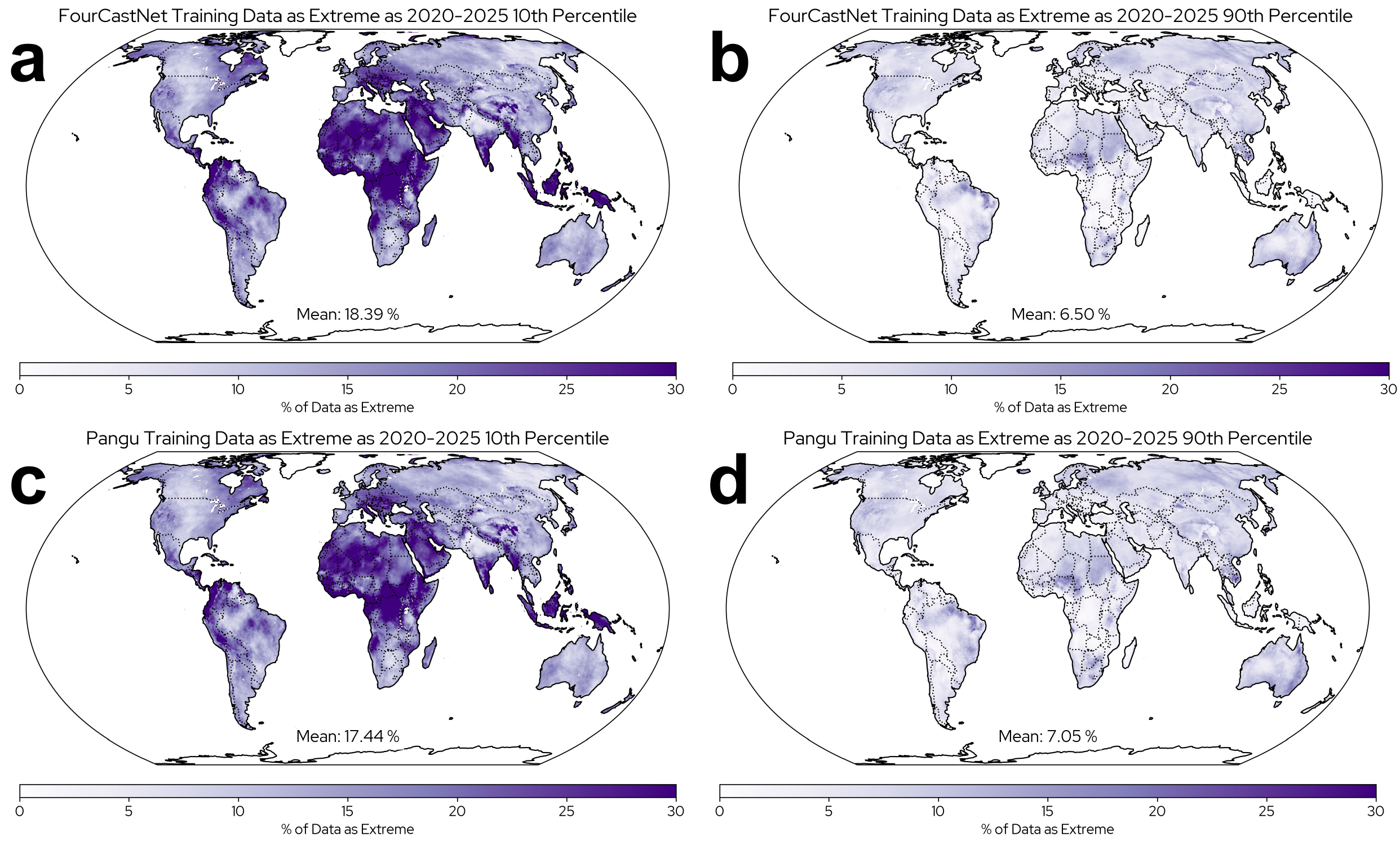}
    \caption{The percent of training data as or more extreme than the 10th and 90th percentiles of 2020-2025 ERA5 temperatures for FourCastNet (a, b) and Pangu (c, d). Global means are shown at the bottom of each panel and are similarly displayed in Figure 3f.}
    \label{fig:figureS1_fourcastnet_regions}
\end{figure}

\begin{figure}
    \centering
    \includegraphics[width=0.99\textwidth]{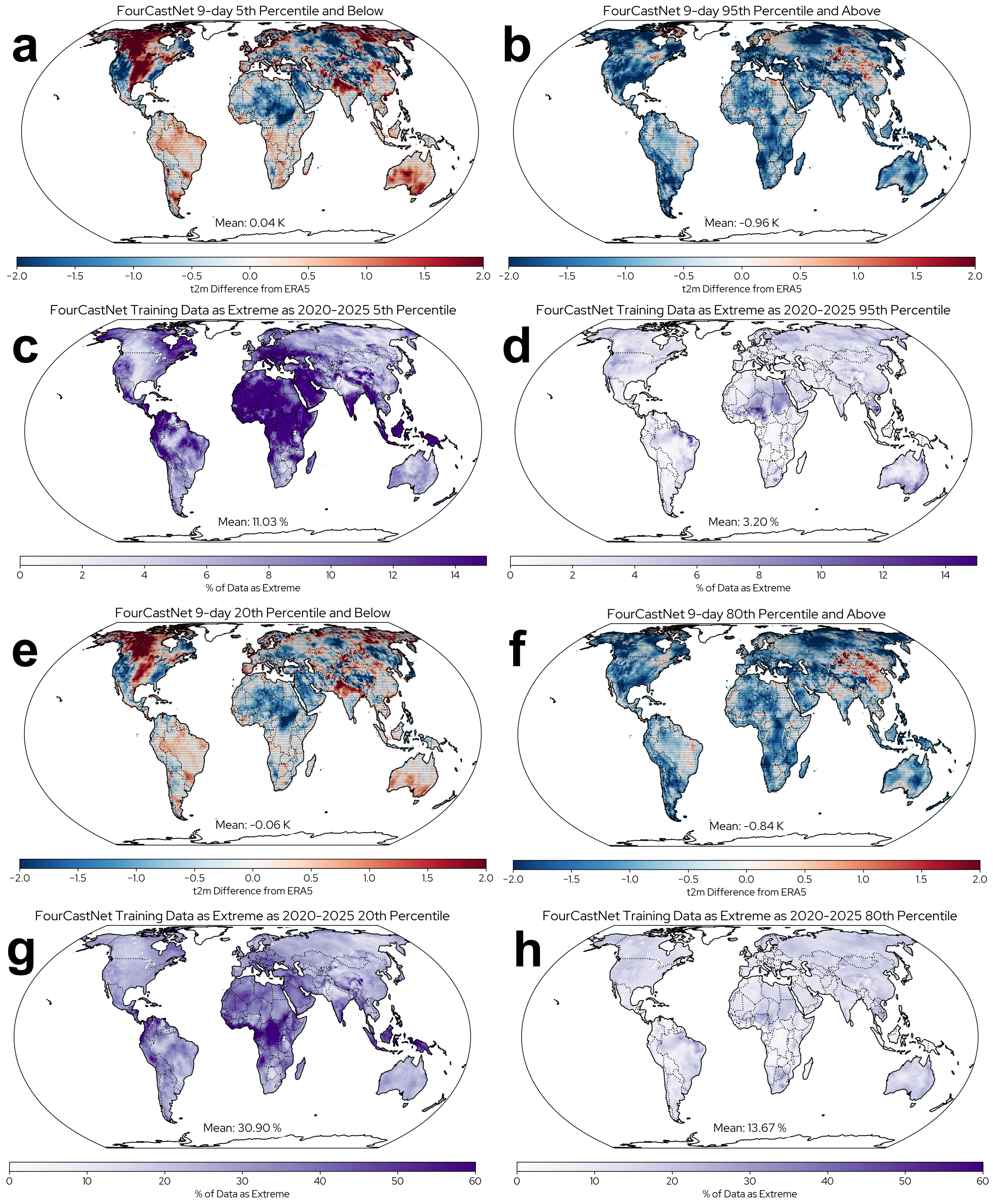}
    \caption{As in Figures 3 (a-d) and \ref{fig:figureS1_fourcastnet_regions}, but for the 5th and 95th (a, c and b, d) and 20th and 80th (e, g and f, h) percentiles of FourCastNet's 9-day lead predictions. Global means are shown at the bottom of each panel. We see similar behavior as in Figure 3, with the hottest percentiles exhibiting a stronger cold bias than the coldest percentiles, in line with their being less training data for hot extremes.\add{ Stippling in a, b, e, and f indicates grid points where the bias is significantly non-zero.}}
    \label{fig:figureS2_fourcastnet_other_percentiles}
\end{figure}

\begin{figure}[p]
    \centering
    \includegraphics[width=0.9\textwidth]{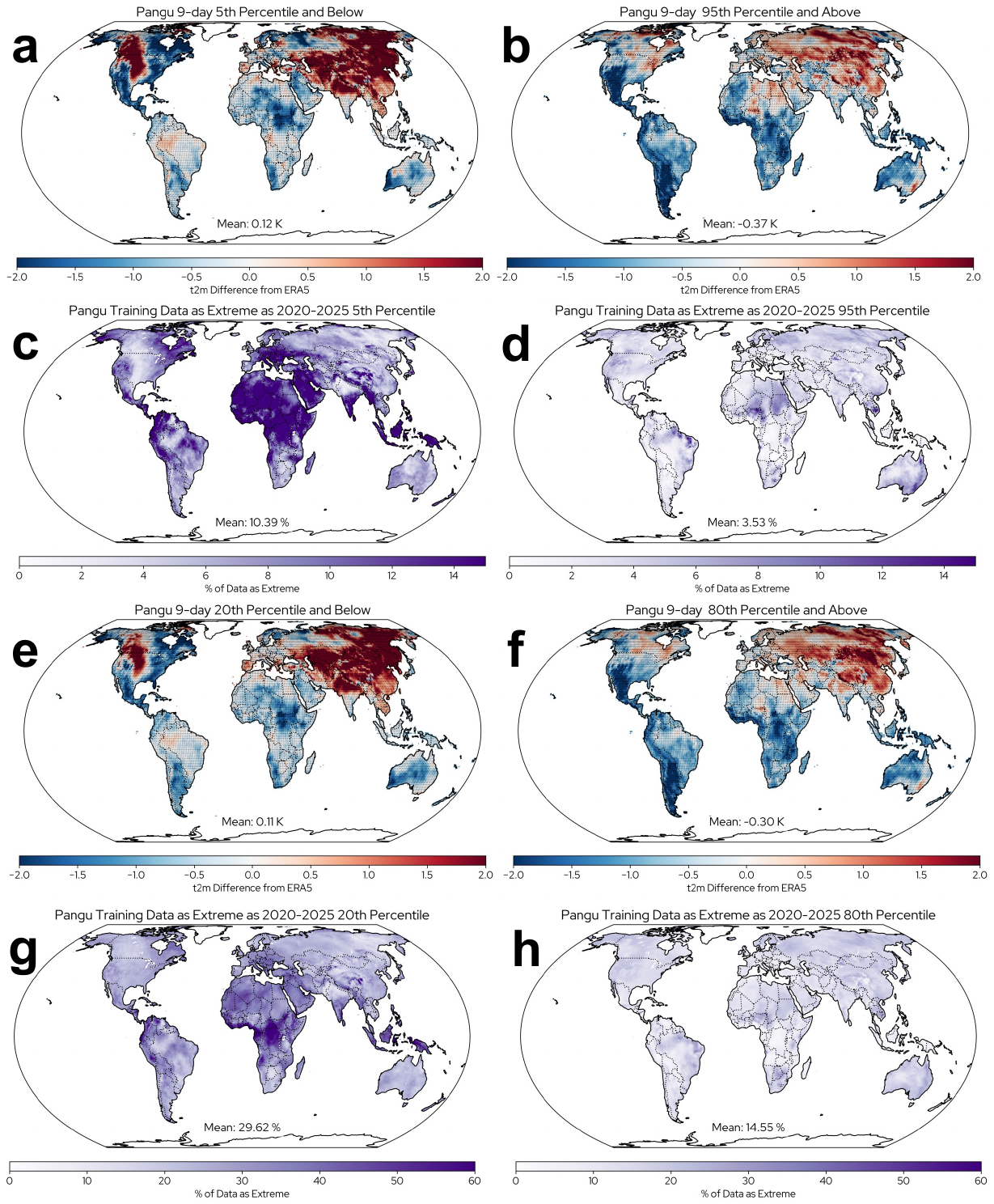}
    \caption{As in Figures 3 (a-d) and \ref{fig:figureS1_fourcastnet_regions}, but for the 5th and 95th (a, c and b, d) and 20th and 80th (e, g and f, h) percentiles of Pangu's 9-day lead predictions. Global means are shown at the bottom of each panel. We see similar behavior as in Figure 3, with the hottest percentiles exhibiting a stronger cold bias than the coldest percentiles, in line with their being less training data for hot extremes.\add{ Stippling in a, b, e, and f indicates grid points where the bias  is significantly non-zero.}}
    \label{fig:figureS3_pangu_other_percentiles}
\end{figure}

\begin{figure}
    \centering
    \includegraphics[width=0.99\textwidth]{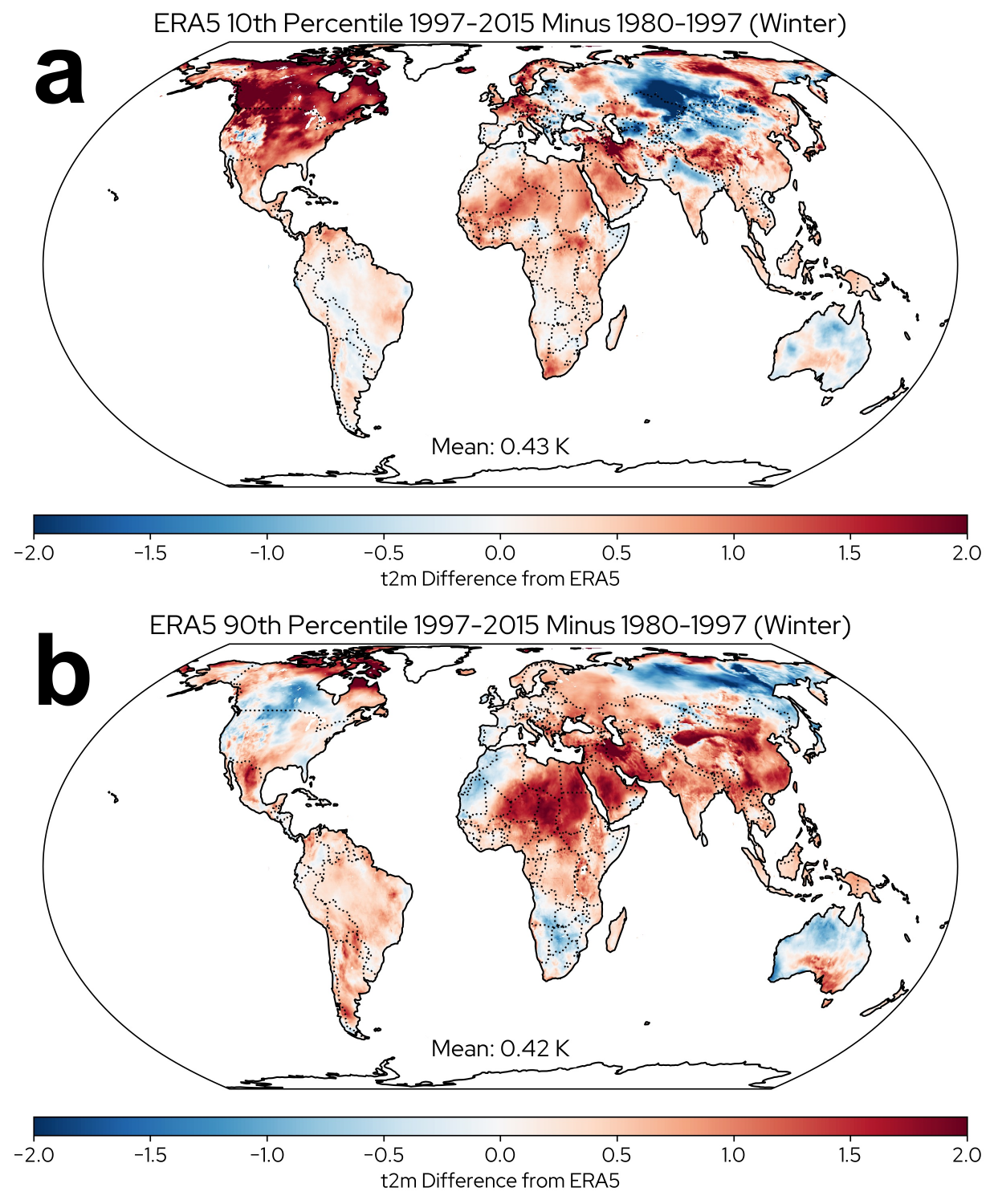}
    \caption{Difference in 10th (a) and 90th (b) percentile ERA5 winter temperatures between 1980-1997 and 1997-2015. We choose these dates as they approximately split the weather model's training sets in half. We see little global mean difference in warming between the 10th and 90th percentiles, unlike what we saw when assessing ACE2's training set (Figure 5b), which showed the change between 1940-1979 and 1980-2022.}
    \label{fig:figureS4_pangu_other_percentiles}
\end{figure}

\clearpage